# A NOVEL APPROACH FOR DISCOVERY MULTI LEVEL FUZZY ASSOCIATION RULE MINING

Pratima Gautam
Department of computer Applications
MANIT, Bhopal

K. R. Pardasani
Department of Mathematics
MANIT, Bhopal

**Abstract**— Finding multilevel association rules in transaction databases is most commonly seen in is widely used in data mining. In this paper, we present a model of mining multilevel association rules which satisfies the different minimum support at each level, we have employed fuzzy set concepts, multi-level taxonomy and different minimum supports to find fuzzy multilevel association rules in a given transaction data set. Apriori property is used in model to prune the item sets. The proposed model adopts a top-down progressively deepening approach to derive large itemsets. This approach incorporates fuzzy boundaries instead of sharp boundary intervals. An example is also given to demonstrate and support that the proposed mining algorithm can derive the multiple-level association rules under different supports in a simple and effective manner.

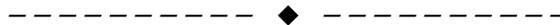

## 1 INTRODUCTION

Association rules are discovered from database tables that store sets of items. Consider a supermarket database where the set of items purchased by a customer on a single visit to a store is recorded as a transaction. The supermarket managers might be interested in finding associations among the items purchased together in one transaction [10] [13]. An example of a supermarket database and a set of association rules derived from the database. The example discovered rule: bread ∧butter ∧milk →apples states that a customer who purchases bread, butter and milk, probably also purchases apples. We refer to the left hand side of the rule as body, and to the right-hand side as head. We also say that the rule is satisfied by a given item set (item set satisfies the rule) if $X \cup Y$ is contained in the set [9], [15]. We say, that the rule is violated by a given item set (item set violates the rule) if the set contains $X$, but does not contain $Y$. Each rule has two measures of its statistical importance and strength: support and confidence. The support of the rule is the number of item sets that satisfy the rule divided by the number of all item sets. The rule confidence is the number of item sets that satisfy the rule divided by the number of item sets that contain $X$. Some approaches for constructing a concept hierarchy and then trying to discover knowledge in the multi-level abstractions to solve this problem are reported in the literature [1] [7] like Apriori algorithm [8], [14] etc. Under the same minimum support and minimum confidence thresholds. This method is simple, but may lead to some undesirable results. Infact different levels should have different support to extract appropriate patterns. Higher support usually exists at higher levels and if one wants to find interesting rules at lower levels, he/she must define lower minimum support values [4], [6].Another trend to deal with the problem is based on fuzzy theory introduced by Zadeh [2], [3], Fuzzy set theory has been used more and more frequently in intelligent systems because of its simplicity and similarity to human reasoning [4]. The use of fuzzy sets to describe association between data extends the types of relationships that may be represented, facilitates the interpretation of rules in linguistic terms, and avoids unnatural boundaries in the partitioning of the attribute domains [11], [12]. Here efficient model based on fuzzy sets and Han's mining approach for multiple-level items is proposed. The proposed model adopts a top-down progressively deepening approach to finding large itemsets [5]. It integrates fuzzy-set concepts, data-mining technologies and multiple-level taxonomy to find fuzzy association rules in given transaction data sets.

## 2. Apriori Algorithm:

The key of mining association rules is to set an appropriate support and confidence values to find frequent itemset. The well-known algorithm, Apriori, exploits the following property: If an itemset is frequent, so are all its subsets [13]. Apriori employs an iterative approach known as level wise search, where $k$-itemsets are used to explore $k+1$-itemsets. First, the set of frequent 1-itemsets is found. This is denoted as $L_1$. $L_1$ is used to find $L_2$, the frequent 2-itemsets, which is used to find $L_3$, and so on, until no more frequent $k$-itemsets can be found. The finding of each $L_k$ requires one full scan of the database. Throughout the level-wise generation of frequent itemsets, an important anti-monotone heuristic is being used to reduce the search space [15].

## 3. Multilevel Association Rule:

Mining association rules at multiple concept levels may, however, lead to discovery of more general and important knowledge from data. Relevant item taxonomies are usually predefined in real-world applications and can be represented as hierarchy trees. Terminal nodes on the trees represent actual items appearing in transactions; internal nodes represent classes or concepts formed from lower-level [4]. For example, the root node ''Food'' is at level 0, the internal nodes representing categories (such as ''Milk'') are at level 1, the internal nodes representing flavor (such as ''Plain Milk'') are at level 2, and the terminal nodes representing choice and brands (such as ''With Sugar'', "Amul") are at level 3, 4. Only terminal nodes appear in transactions. Nodes in predefined taxonomies are first encoded using sequences of numbers and the symbol ''*'' according to their positions in the hierarchy tree. For example, the internal node ''Milk'' in fig.1 is represented by 1***, the internal node ''Plain Milk'' by 11*, and the terminal node ''with sugar plain milk'' by 111*, Amul with sugar plain milk by 1111. A top-down progressively deepening search approach is used and exploration of ''level-crossing'' association relationships is allowed. Candidate itemsets at certain levels may thus contain items at lower levels [6] [14]. For example, candidate 2-itemsets at level 2 are not limited to containing only pairs of large items at level 2. Instead, large items at level 2 may be paired with large items at level 1 to form candidate 2-itemsets at level 2 (such as {11*, 2**}

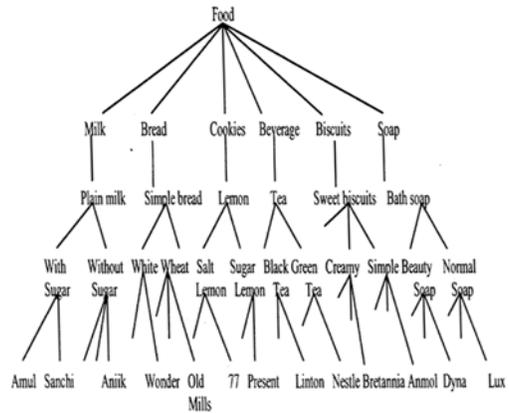

Fig. 1. The taxonomy for the relevant data items

## 3 The Proposed Model:

Apriori algorithm [3] to examine data items at multiple levels of abstraction under the same minimum support and confidence thresholds. This method is simple, but some undesirable results may come out. First, different level has different support. Larger support usually exists at higher levels and if one wants to find interesting rules at lower levels. Some algorithms have been developed for efficient mining. One approach progressively reduced the minimum support thresholds at lower levels of abstraction .The underlying assumption in [1], [4] was only to explore the descendants of the frequent items since if an item rarely occurs its descendants will occur even less frequently and, thus, are uninteresting. In this paper, we use Han and Fu's encoding scheme, as described in [5], [7] to represent nodes in predefined taxonomies for mining multilevel rules. The proposed algorithm finds all the large itemsets for the given transactions by comparing the fuzzy count of each candidate itemset with its support threshold. Furthermore, some pruning strategies are used to reduce the number of candidate S itemsets generated. The proposed algorithm is given in the following steps:

Step 1:
Encode taxonomy using a sequence of numbers and the symbol ''*'', with the $l$th number representing the branch number of a certain item at levels.

Step-2:
Determine $\chi \in \{2, 3, 4....\}$ (Maximum item threshold). $\chi$ is a threshold to determine maximum number of items in a transaction by which the transaction may or may not be considered in the process of generating rules mining. In this case, the process just considers all transactions with the number of items in the transactions less than or equal to $\chi$. Formally, let D be a universal set of transactions. $M \subseteq D$ is considered as a subset of qualified transactions for generating rules mining that the number of items in its transactions is not greater than $\chi$ as defined by:

$$M = \{T \mid card(T) \leq \chi, T \in D\}, \quad (1)$$

Where card (T) is the number of items in transaction T.

Step-3:
Set k = 1, where k is used to store the level number being processed whereas k ∈ {1, 2, 3, 4} (as we consider up to 4-levels of hierarchies).

Step-4: Set q=1, where q is an index variable to determine the number of combination of items in itemsets called q-itemsets. q ∈ {1, 2, 3, 4} (as we consider up to 4-itemsets at each level of hierarchy).

Step-5
Determine minimum support for q-itemsets at level k, denoted by $\beta_q^k \in (0, |M|]$ as a minimum threshold of a combination items appearing in the whole qualified transactions, where |M| is the number of qualified transactions. $\beta_q^k$ may have same value for every q at level k.

Step-6
Group the items with the same first k digits in each transaction $T_j$, and add the occurrence of the items in the same groups in $T_i$. Denote the amount of the j-th group $I_j^q$ for $T_i$ as $v_{ij}^q$.

Step 7:
Construct every candidate q-itemset, $I^q$ as a fuzzy set on set of qualified transactions, M. A fuzzy membership function $\mu$ is a mapping:
$\mu_{I^q} : M \to [0, 1]$ as defined by:

$$\mu_{I^q}(T) = v_{ij}^q \cdot \inf_{i \in I^q}\left\{\frac{\eta_T(i)}{Card(T)}\right\}, \forall T \in M \quad (2)$$

Where T be a qualified transaction in which T can be regarded also as a subset of items. $T \subseteq D$
A Boolean membership function, η, is a mapping

$$\eta_T : D \to \{0, 1\}$$

as defined by:
$$\eta_T(i) = \begin{cases} 1, i \in T \\ 0, otherwise \end{cases} \quad (3)$$

Such that if an item, i, is an element of T then $\eta_T(i) = 1$, otherwise $\eta_T(i)=0$

Step - 8:
Calculate *support* for every (candidate) q-itemset using the following equations:
$$Support(I^q)^k = \sum_{T \in M} v_{ij}^q \cdot \mu_{I^q}^k \quad (4)$$

M is the set of qualified transactions; it can be proved that (4) satisfied the following property:
$$\sum_{i \in D} Support(i) = |M| \quad (5)$$

For $q=1$, $I^q$ can be considered as a single item. if q>1 then generate candidate set $C_2^k$ has to following steps for each newly from 2-itemsets.

Step-9:
$I^q$ Will be stored in the set of frequent q-itemsets, $N_q^k$ if and only if support ($I_j^q$) $\geq \beta_q^k$

Step-10
Set q = q+1, for the same level k and if q >4, then go to Step-12.

Step-11
Looking for possible/candidate q-itemsets from $L_{q-1}$ by the following rules:

A q-itemset $I^q$ will be considered as a candidate q-itemset if $I^q$ satisfies:
$$\forall F \subset I^p \mid F \mid = K - 1 \Rightarrow F \in N_{q-1}$$

For example, $I^q$ = {1112, 2112, 3122, 3122, 4112, 4112} will be considered as a candidate 6-itemset, iff: {1112, 2112, 3122, 2113, 1113}, {1112, 3112, 4112, 4112, 5123, 5123}, {1113, 1113, 2123, 3112, 3112, 6133} and {3122, 4122, 5123, 5123, 6133} in $N_4$. If there is no candidate q-itemset then go to Step-12. Otherwise, the process is going to Step-4.

Step-12
Similar to Apriori Algorithm, confidence of an association rule mining, A ⇒ B, can be calculated by the following equation:

$$Conf(A \Rightarrow B) = P(B \mid A) = \frac{Support(A \cup B)}{Support(A)} \quad (6)$$

Where A, B ∈ D
It can be followed that (5) can also be expressed by

$$Conf(A \Rightarrow B) = \frac{\sum_{T \in M} \inf_{i \in A \cup B}(v_{ij}^q \cdot \mu_i(T))}{\sum_{T \in M} \inf_{i \in A}(v_{ij}^q \cdot \mu_i(T))}$$

Where A and B is any q-itemsets in Lq. $[\mu_i(T) = \mu_{(i)}(T)]$. Therefore, support of an itemset as given by (4) can be expressed as following:

$$Support(I^q)^k = \sum_{T \in M} \inf_{i \in I^q}(\mu_i(T)) \quad (7)$$

Step -13
Set k=k+1 and go to step-6 (for repeating the whole processing for next level).

## 4. AN ILLUSTRATIVE EXAMPLE

An illustrative example is given to understand well the concept of the proposed model and algorithm and how the process of the generating fuzzy association rule mining is performed step by step. The process is started from a given transactional database as shown in Table 1[a].

TABLE 1[a]

| Trans_ID | List of items |
|---|---|
| T1 | 1112, 2112, 3122, 2113, 1113 |
| T2 | 1112, 2112, 3122, 3122, 4112, 4112 |
| T3 | 1112, 3112, 4112, 4112, 5123, 5123 |
| T4 | 1111, 2112, 2112, 4112, 5123, 5123, 6123, 6123 |
| T5 | 1113, 1113, 2123, 3112, 3112, 6133 |
| T6 | 3122, 4122, 5123, 5123, 6133 |
| T7 | 4112, 4112, 5123, 5123, 5123, 6123, 6123 |
| T8 | 1113, 2112, 3122, 6123, 1113 |
| T9 | 1111, 6123, 5134, 6123, 5134 |
| T10 | 6123, 4112, 6123, 6123, 4112 |
| T11 | 1123, 4112, 1123 |
| T12 | 1111, 5134, 5134 |
| T13 | 3112, 1123, 3112, 3112 |

**Table1 [b]**
**Codes of item name**

| Item name (terminal node) | Code | Item name (internal node) | Code |
|---|---|---|---|
| Amul with sugar plain milk | 1111 | Milk | 1*** |
| Sanchi with sugar plain milk | 1112 | Bread | 2*** |
| Anik without sugar plain milk | 1123 | Cookies | 3*** |
| Wonder white simple bread bread | 2112 | Beverage | 4*** |
| Old Mills wheat simple bread bread | 2123 | Tooth paste | 5*** |
| 77salt lemon lemon cookies | 3112 | Soap | 6*** |
| Present sugar lemon lemon cookies | 3122 | Plain milk | 11** |
| Limton black tea tea beverage | 4112 | Plain bread bread | 21** |
| Nestle green tea tea beverage | 4122 | Lemon cookies | 31** |
| Britannia creamy sweet biscuits biscuits | 5123 | Tea beverage | 41** |
| Anmol simple sweet biscuits | 5134 | Sweet biscuits biscuits | 51** |
| Dyna cream soap bath soap soap | 6123 | Bath soap soap | 61** |
| With sugar plain milk | 111* | | |
| White plain bread bread | 211* | | |
| Salt lemon lemon cookies | 311* | | |
| Without sugar plain milk | 112* | | |
| Wheat plain bread bread | 212* | | |
| Sugar lemon lemon cookie | 312* | | |

**Step-1:**
We are using encoded transaction table1. For example, the item '' foremore plain milk '' in figure1 is encoded as '122', in which the first digit '1' represents the code 'milk' at level 1, the second digit '2' represents. The flavor 'plain (milk)' at level 2, and the third digit '2' represents the brand 'foremore' at level 3.

**Step-2**
Suppose that $\chi$ arbitrarily equals to 6; that means qualified transaction is regarded as a transaction with no more than 6 items purchased in the transaction. Result of this step is a set of qualified transaction as seen in Table 2. Where M={T1,T2,T3, T5,T6 ,T7 ,T8, ,T10,T11,T12, T13 }

**Table [2]**

| Transaction id | list of items |
|---|---|
| T1 | 1112, 2112, 3122, 2113, 1113 |
| T2 | 1112, 2112, 3122, 3122, 4112, 4112 |
| T3 | 1112, 3112, 4112, 4112, 5123, 5123 |
| T5 | 1113, 1113, 2123, 3112, 3112, 6133 |
| T6 | 3122, 4122, 5123, 5123, 6133 |
| T8 | 1113, 2112, 3122, 6123, 1113 |
| T9 | 1111, 6123, 5134, 6123, 5134 |
| T10 | 6123, 4112, 6123, 6123, 4112 |
| T11 | 1123, 4112, 1123 |
| T12 | 1111, 5134, 5134 |
| T13 | 3112, 1123, 3112, 3112 |

The Qualified set of Transaction (M)

**Step-3:**
Let k $\in$ {1, 2, 3, 4}, where k is used to store the level number being processed.

**Step-4:**
The process is started by looking for support of 1-itemsets (q =1) for level k = 1.

**Step-5:**
Since q $\in$ {1, 2, 3, 4}. It is arbitrarily given $\beta_q^1$ = 0.36, $\beta_q^2$ = 0.3, $\beta_q^3$ = 0.2, $\beta_q^4$ = 0.16. That means the system just considers support of q-itemsets that is greater than or equal to 0.36, for K=1, and greater than or equal to 0.3, for K=2, and greater than or equal to 0.2, for K=3 and greater than or equal to 0.16, for k = 4.

**Step-6:**
All the items in the transactions are first grouped at level one and their corresponding occurrence are added. Take the items in transaction T1 as an example. The items (2112) and (2113) grouped (2***, 2), representation of grouped data items with their occurrence count is given in table3.

**Table-3**

| Trans_ID | List of item |
|---|---|
| T1 | (1***, 1) (2***, 2) (3***, 1) |
| T2 | (1***, 1) (2***, 1) (3***, 2) (4***, 2) |
| T3 | (1***, 1) (3***, 1) (4***, 2) (5***, 2) |
| T5 | (1***, 2) (2***, 1) (3***, 2) (6***, 1) |
| T6 | (3***, 1) (5***, 2) (4***, 1) (6***, 1) |
| T8 | (1***, 2) (2***, 1) (3***, 1) (6***, 1) |
| T9 | (1***, 1) (6***, 2) (5***, 2) |
| T10 | (6***, 3) (4***, 2) |
| T11 | (1***, 2) (4***, 1) |
| T12 | (1***, 1) (5***, 2) |
| T13 | (3***, 3) (1***, 1) |

The Grouped items in transaction

**Step-7**
Every q-itemset $I^q$ is represented as a fuzzy set on set of qualified transactions M as given by the following results:

**Level k = 1**
**Min_support ($\beta_q^1$) = 0.36**

| 1-Itemset | Support of 1-Itemsets |
|---|---|
| {1***} = {0.4/T1, 0.16/T2, 0.16/T3, 0.33/T5, 0.4/T8, 0.2/T9, 0.66/T11, 0.33/T12, 0.25/T13} | {1***} = 2.89 |
| {2***} = {0.4/T1, 0.16/T2, 0.16/T5, 0.2/T8} | {2***} = 0.92 |
| {3***} = {0.2/T1, 0.33/T2, 0.16/T3, 0.33/T5, 0.2/T6, 0.2/T8, 0.75/T13} | {3***} = 2.17 |
| {4***} = {0.33/T2, 0.33/T3, 0.2/T6, 0.4/T10, 0.33/T11} | {4***} = 1.59 |
| {5***} = {0.33/T3, 0.4/T6, 0.4/T9, 0.66/T12} | {5***} = 1.79 |
| {6***} = {0.16/T5, 0.2/T6, 0.2/T8, 0.4/T9, 0.6/T10} | {6***} = 1.56 |

**Table- 4**

$$N_1^1$$

| 1-itemset | Min_Support |
|---|---|
| {1***} | 2.89 |
| {2***} | 0.92 |
| {3***} | 2.17 |
| {4***} | 1.59 |
| {5***} | 1.79 |
| {6***} | 1.59 |

All 1-items considered for further process because their support $>= \beta_1^1$

**2-Itemset** **Support of 2-Itemsets**
{1***, 2***} = {0.4/T1 $\wedge$ 0.4/T1, 0.16/T2 $\wedge$ 0.16/T2, 0.33/T5 $\wedge$ 0.16/T5, 0.4/T8 $\wedge$ 0.2/T8}     {1***, 2***} = 0.92
   = {0.4/T1, 0.16/T2, 0.16/T5, 0.2/T8}
{1***, 3***} = {0.4/T1 $\wedge$ 0.2/T1, 0.16/T2 $\wedge$ 0.33/T2,     {1***, 3***} = 1.3

0.16/T3 ∧ 0.16/T3, 0.33/T5 ∧ 0.33/T5
0.4/T8 ∧ 0.2/T8, 0.25/T13 ∧ 0.75/T13}
= {0.2/T1, 0.16/T2, 0.16/T3, 0.33/T5, 0.2/T8, 0.25/T13}

{1***, 4***} = {0.16/T2 ∧ 0.33/T2, 0.16/T3 ∧ 0.33/T3, 0.66/T11 ∧ 0.33/T11}    {1***, 4***} = 0.65
= {0.16/T2, 0.16/T3, 0.33/T11}

{1***, 5***} = {0.16/T3 ∧ 0.33/T3, 0.2/T9 ∧ 0.4/T9, 0.33/T12 ∧ 0.66/T12}    {1***, 5***} = 0.69
= {0.16/T3, 0.2/T9, 0.33/T12}

{1***, 6***} = {0.33/T5 ∧ 0.16/T5, 0.4/T8 ∧ 0.2/T8, 0.2/T9 ∧ 0.4 T9}    {1***, 6***} = 0.56
= {0.16/T5, 0.2/T8, 0.2/T9}

{2***, 3***} = {0.4/T1 ∧ 0. 2/T1, 0.16/T2 ∧ 0.33/T2, 0.16/T5 ∧ 0.33/T5, 0.2/T8 ∧ 0.2/T8}    {2***, 3***} = 0.72
= {0.2/T1, 0.16/T2, 0.16/T5, 0.2/T8}

{2***, 4***} = {0.16/T2 ∧ 0.33/T2}    {2***, 4***} = 0.16
= {0.16/T2}

{2***, 5***} = { }    {2***, 5***} = { }

{2***, 6***} = {0.16/T5 ∧ 0.16/T5, 0.2/T8 ∧ 0.2/T8}    {2***, 6***} = 0.36
= {0.16/T5, 0.2/T8}

{3***, 4***} = {0.33/T2 ∧ 0.33/T2, 0.16/T3 ∧ 0.33/T3, 0.2/T6 ∧ 0.2/T6}    {3***, 4***} = 0.69
= {0.33/T2, 0.16/T3, 0.2/T6}

{3***, 5***} = {0.16/T3 ∧ 0.33/T3, 0.2/T6 ∧ 0.4/T6}    {3***, 5***} = 0.36
= {0.16/T3, 0.2/T6}

{3***, 6***} = {0.33/T5 ∧ 0.16/T5, 0.2/T6 ∧ 0.2/T6, 0.2/T8 ∧ 0.2/T8}    {3***, 6***} = 0.56
= {0.16/T5, 0.2/T6, 0.2/T8}

{4***, 5***} = {0.33/T3 ∧ 0.33/T3, 0.2/T6 ∧ 0.4/T6}    {4***, 5***} = 0.53
= {0.33/T3, 0.2/T6}

{4***, 6***} = {0.2/T6 ∧ 0.2/T6, 0.4/T10 ∧ 0.6/T10}    {4***, 6***} = 0.6
= {0.2/T6, 0.4/T10}

{5***, 6***} = {0.4/T6 ∧ 0.2/T6, 0.4/T9 ∧ 0.4/T9}    {5***, 6***} = 0.6
= {0.2/T6, 0.4/T9}

**Table- 5**

$$N_2^1$$

| 2-itemset | Min_Support |
|---|---|
| {1***, 2***} | 0.92 |
| {1***, 3***} | 1.8 |
| {1***, 4***} | 0.65 |
| {1***, 5***} | 0.69 |
| {1***, 6***} | 0.56 |
| {2***, 3***} | 0.72 |
| {2***, 4***} | 0.16 |
| {2***, 5***} | { } |
| {2***, 6***} | 0.36 |
| {3***, 4***} | 0.69 |
| {3***, 5***} | 0.36 |
| {3***, 6***} | 0.56 |
| {4***, 5***} | 0.53 |
| {4***, 6***} | 0.6 |
| {5***, 6***} | 0.6 |

The {2***, 4***}, {2***, 5***} cannot be considered for further process because their *support* >= $\beta_2^1$

**3-Itemset**                                   **Support of 3- Itemsets**

{1***, 2***, 3***} = {0.4/T1 ∧ 0.4/T1 ∧ 0.2/T1, 0.16/T2 ∧ 0.16/T2 ∧ 0.33/T2, 0.33/T5 ∧ 0.16/T5 ∧ 0.33/T5, 0.4/T8 ∧ 0.2/T8 ∧ 0.2/T8}    {1***, 2***, 3***} = 0.72
= {0.2/T1, 0.16/T2, 0.16/T5, 0.2/T8}

{1***, 2***, 4***} = {0.16/T2 ∧ 0.16/T2 ∧ 0.33/T2}    {1***, 2***, 4***} = 0.16
= {0.16/T2}

{1***, 2***, 6***} = {0.33/T5 ∧ 0.16/T5 ∧ 0.16/T5, 0.4/T8 ∧ 0.2/T8 ∧ 0.2/T8}    {1***, 2***, 6***} = 0.36
= {0.16/T5, 0.2/T8}

{1***, 3***, 4***} = {0.16/T2 ∧ 0.33/T2 ∧ 0.33/T2, 0.16/T3 ∧ 0.16/T3 ∧ 0.33/T3}    {1***, 3***, 4***} = 0.32
= {0.16/T2, 0.16/T3}

{1***, 3***, 5***} = {0.33/T5 ∧ 0.33/T5 ∧ 0.16/T5} = {0.16/T5}    {1***, 3***, 5***} = 0.16

{1***, 3***, 6***} = {0.33/T5 ∧ 0.33/T5 ∧ 0.16/T5, 0.4/T8 ∧ 0.2/T8 ∧ 0.33/T8}    {1***, 3***, 6***} = 0.36
= {0.16/T5, 0.2/T8}

{1***, 4***, 5***} = {0.16/T3 ∧ 0.33/T3 ∧ 0.33/T3} = {0.16/T3}    {1***, 4***, 5***} = 0.16

{1***, 4***, 6***} = { }    {1***, 4***, 6***} = { }

{2***, 3***, 6***} = {0.16/T5 ∧ 0.33/T5 ∧ 0.16/T5, 0.2/T8 ∧ 0.2/T8 ∧ 0.2/T8}    {2***, 3***, 6***} = 0.36
= {0.16/T5, 0.2/T8}

{2***, 4***, 6***} = { }    {2***, 4***, 6***} = { }

**Table-6**

$$N_3^1$$

| 3-itemset | Support |
|---|---|
| {1***, 2***, 3***} | 0.72 |
| {1***, 2***, 6***} | 0.36 |
| {1***, 3***, 6***} | 0.36 |
| {2***, 3***, 6***} | 0.36 |
| {3***, 4***, 5***} | 0.36 |

The {1***, 2***,4***}, {1***, 3***, 4***},{1***, 3***, 5***},{1***, 4***, 5***}, {1***, 4***, 6***}, {2***, 3***, 4***}, {2***, 4***, 6***}, {3***, 4***, 6***}, {4***, 5***, 6***}  cannot be considered for further process because their *support* <= $\beta_3^1$.

**4-Itemset**

{1***, 2***, 3***, 6***} = {0.33/T5 ∧ 0.16/T5 ∧ 0.33/T5 ∧ 0.16/T5, 0.4/T8 ∧ 0.2/T8 ∧ 0.2/T8 ∧ 0.2/T8} = 0.16/T5, 0.2/T8

**Support of 4- Itemsets** = 0.36

**Table-7**

$$N_4^1$$

| 4-itemset | Support |
|---|---|
| {1***, 2***, 3***, 6***} | 0.36 |

All 4-items considered for further process because their *support* >= $\beta_4^1$.

**Level k = 2**
Min_support ($\beta_q^2$) = 0.3

**1-Itemset**                                       **Support of 1- Itemsets**

{11**} = {0.4/T1, 0.16/T2, 0.16/T3, 0.33/T5, 0.4/T8, 0.2/T9, 0.66/T11, 0.33/T12, 0.25/T13}    {11**} = 2.89

{21**} = {0.4/T1, 0.16/T2, 0.16/T5, 0.2/T8}    {21**} = 0. 92

{31**} = {0.2/T1, 0.33/T2, 0.16/T3, 0.33/T5, 0.2/T6, 0.2/T8, 0.75/T13}    {31**} = 2.17

{61**} = {0.16/T5, 0.2/T6, 0.2/T8, 0.4/T9, 0.6/T10}    {61**} = 1.56

**Table- 8**

$$N_1^2$$

| 1-itemset | Min_Support |
|---|---|
| {11**} | 2.89 |
| {21**} | 0.92 |
| {31**} | 2.17 |
| {61**} | 1.56 |

All 1-items considered for further process because their *support* >= $\beta_1^2$.

**2-Itemset**  **Support of 2-Itemsets**
{11**, 21**} = {0.4/T1 ∧ 0.4/T1, 0.16/T2       {11**, 21**} = 0.92
∧ 0.16/T2, 0.33/T5 ∧ 0.16/T5, 0.4/T8 ∧ 0.2/T8}
= {0.4/T1, 0.16/T2, 0.16/T5, 0.2/T8}
{11**, 31**} = {0.4/T1 ∧ 0.2/T1, 0.16/T2 ∧ 0.33/T2,   {11**, 31**} = 1.3
0.16/T3 ∧ 0.16/T3, 0.33/T5 ∧ 0.33/T5 0.4/T8 ∧
0.2/T8, 0.25/T13 ∧ 0.75/T13}
= {0.2/T1, 0.16/T2, 0.16/T3, 0.33/T5, 0.2/T8, 0.75/T13}
{11**, 61**} = {0.33/T5 ∧ 0.16/T5, 0.4/T8 ∧ 0.2/T8,   {11**, 61**} = 0.56
0.2/T9 ∧ 0.4 T9}   = {0.16/T5, 0.2/T8, 0.2/T9}
{21**, 31**} = {0.4/T1 ∧ 0.2/T1, 0.16/T2 ∧ 0.33/T2,   {21**, 31**} = 0.72
0.16/T5 ∧ 0.33/T5, 0.2/T8 ∧ 0.2/T8}
= {0.2/T1, 0.16/T2, 0.16/T5, 0.2/T8}
{21**, 61**} = {0.16/T5 ∧ 0.16/T5, 0.2/T8 ∧ 0.2/T8}   {21**, 61**} = 0.36
= {0.16/T5, 0.2/T8}
{31**, 61**} = {0.33/T5 ∧ 0.16/T5, 0.2/T6 ∧ 0.2/T6,   {31**, 61**} = 0.56
0.2/T8 ∧ 0.2/T8}   = {0.16/T5, 0.2/T6, 0.2/T8}

**Table- 9**

$$N_2^2$$

| 2-itemset | Min_Support |
|---|---|
| {11**, 21**} | 0.92 |
| {11**, 31**} | 1.3 |
| {11**, 61**} | 0.56 |
| {21**, 31**} | 0.72 |
| {21**, 61**} | 0.36 |
| {31**, 61**} | 0.56 |

All 2-items considered for further process because their *support* >= $\beta_2^2$.

**3-Itemset**  **Support of 3-Itemsets**
{11**, 21**, 31**} = {0.4/T1 ∧ 0.4/T1 ∧ 0.2/T1,   {11**, 21**, 31**} = 0.72
0.16/T2 ∧ 0.16/T2 ∧ 0.33/T2, 0.33/T5 ∧
0.16/T5 ∧ 0.33/T5, 0.4/T8 ∧ 0.2/T8 ∧ 0.2/T8}
= {0.2/T1, 0.16/T2, 0.16/T5, 0.2/T8}
{11**, 21**, 61**} = {0.33/T5 ∧ 0.16/T5 ∧         {11**, 21**, 61**} = 0.36
0.16/T5, 0.4/T8 ∧ 0.2/T8 ∧ 0.2/T8}
= {0.16/T5, 0.2/T8}
{21**, 31**, 61**} = {0.16/T5 ∧ 0.33/T5 ∧        {21**, 31**, 61**} = 0.36
0.16/T5, 0.2/T8 ∧ 0.2/T8 ∧ 0.2/T8}
= {0.16/T5, 0.2/T8}

**Table- 10**

$$N_3^2$$

| 3-itemset | Min_Support |
|---|---|
| {11**, 21**, 31**} | 0.72 |
| {11**, 21**, 61**} | 0.36 |
| {21**, 31**, 61**} | 0.36 |

All 3-items considered for further process because their *support* >= $\beta_3^2$.

**4-Itemset**

{11**, 21**, 31**, 61**} = {0.33/T5 ∧ 0.16 /T5 ∧ 0.33/T5 ∧ 0.16/T5,
0.4/T8 ∧ 0.2/T8 ∧ 0.2/T8 ∧ 0.2/T8}
 = 0.16/T5, 0.2/T8

**Support of 4- Itemsets**  {11**, 21**, 31**, 61**} = 0.36

Table-11

$$N_4^2$$

| 4-itemset | Support |
|---|---|
| {11**, 21**, 31**, 61**} | 0.36 |

All 4-items considered for further process because their *support* >= $\beta_4^2$.

**Level k = 3**
Min_support ($\beta_q^3$) = 0.2

**1-Itemset**  **Support of 1-Itemsets**
{111*} = {0.4/T1, 0.16/T2, 0.16/T3, 0.33/T5,   {111*} = 1.98
0.4/T8, 0.2/T9, 0.33/T12}
{112 *} = {0.66/T11, 0.25/T13}   {112*} = 0.9
{211*} = {0.4/T1, 0.16/T2, 0.2/T8}   {211*} = 0.7
{212*} = {0.16/T2}   {212*} = 0.16
{311*} = {0.16/T3, 0.33/T5, 0.75/T13}   {311*} = 1.24
{312*} = {0.2/T1, 0.33/T2, 0.2/T6, 0.2/T8}   {312*} = 0.93
{612*} = {0.2/T8, 0.4/T9, 0.6/T10}   {612*} = 1.2
{613*} = {0.16/T5, 0.2/T6}   {613*} =  0.36

**Table- 12**

$$N_1^3$$

| 1-itemset | Min_Support |
|---|---|
| {111*} | 1.98 |
| {112*} | 0.91 |
| {211*} | 0.76 |
| {212*} | 0.16 |
| {311*} | 1.24 |
| {312*} | 0.93 |
| {612*} | 1.2 |
| {613*} | 0.36 |

{212*} cannot be considered for further process because their *support* <= $\beta_1^3$.

**2-Itemset**  **Support of 2-Itemsets**
{111*, 112*} = { }   {111*, 112*} = { }
{111*, 211*} = {0.4/T1 ∧ 0.4/T1, 0.16/T2   {111*, 211*} = 0.76
∧ 0.16/T2, 0.4/T8 ∧ 0.2/T8}

= {0.4/T1, 0.16/T2, 0.2/T8}
{111*, 311*} = {0.16/T3 $\wedge$ 0.16/T3, 0.33/T5 $\wedge$ 0.33/T5}  = {0.16/T3, 0.33/T5}   {111*, 311*} = 0.49
{111*, 312*} = {0.4/T1 $\wedge$ 0.2/T1, 0.16/T2 $\wedge$ 0.33/T2, 0.4/T8 $\wedge$ 0.2/T8}   {111*, 312*} = 0.56
= {0.2/T1, 0.16/T2, 0.2/T8}
{111*, 612*} = {0.4/T8 $\wedge$ 0.2/T8, 0.2/T9 $\wedge$ 0.4/T9} = {0.2/T8, 0.2/T9}   {111*, 612*} = 0.4
{111*, 613*} = {0.33/T5 $\wedge$ 0.16/T5} = {0.16/T5}   {111*, 613*} = 0.16
{112*, 211*} = { }   {112*, 211*} = { }
{112*, 311*} = {0.25/T13 $\wedge$ 0.75/T13} = {0.25}   {112*, 311*} = 0.25
{112*, 312*} = { }   {112*, 312*} = { }
{112*, 612*} = { }   {112*, 612*} = {
{112*, 613*} = { }   {112*, 613*} = { }
{211*, 212*} = { }   {211*, 212*} = { }
{211*, 311*} = { }   {211*, 311*} = { }
{211*, 312*} = {0.4/T1 $\wedge$ 0.2/T1, 0.16/T2 $\wedge$ 0.33/T2, 0.2/T8 $\wedge$ 0.2/T8}= {0.2/T1, 0.16/T2, 0.2/T8}   {211*, 312*} = 0.56
{211*, 612*} = {0.2/T8 $\wedge$ 0.2/T8} = {0.2}   {211*, 612*} = 0.2
{211*, 613*} = { }   {211*, 613*} = { }
{311*, 312*} = { }   {311*, 312*} = { }
{311*, 612*} = { }   {311*, 612*} = { }
{311*, 613*} = {0.33/T5 $\wedge$ 0.16/T5} = {0.16/T5}   {311*, 613*} = 0.16
0.16{312*, 612*} = {0.2/T8 $\wedge$ 0.2/T8} = {0.2/T8}   {312*, 612*} = 0.2
{312*, 613*} = {0.2/T6 $\wedge$ 0.2/T6} = {0.2/T6}   {312*, 613*} = 0.2

**Table- 13**

$$N_2^3$$

| 2-itemset | Min_Support |
|---|---|
| {111*, 211*} | 0.76 |
| {111*, 311*} | 0.49 |
| {111*, 312*} | 0.56 |
| {111*, 612*} | 0.4 |
| {112*, 311*} | 0.25 |
| {211*, 312*} | 0.56 |
| {211*, 612*} | 0.2 |
| {312*, 612*} | 0.2 |
| {312*, 613*} | 0.2 |

The {111*, 112*}, {111*, 613*}, {112*, 211*}, {112*, 312*},{112*, 612*}, {112*, 613*}, {211*, 311*}, {211*, 613*}, {311*, 312*}, {311*, 612*}, {311*, 613*} cannot be considered for further process because their *support* <= $\beta_2^3$.

**3-Itemset**                                             **Support of 3- Itemsets**

{111*, 211*, 312*} = {0.4/T1 $\wedge$ 0.4/T1 $\wedge$ 0.2/T1, 0.16/T2 $\wedge$ 0.16/T2 $\wedge$ 0.33/T2, 0.4/T8 $\wedge$ 0.2/T8 $\wedge$ 0.2/T8} = {0.2/T1, 0.16/T2, 0.2/T8}   {111*, 211*, 312*} = 0.56
{111*, 211*, 612*} = {0.4/T8 $\wedge$ 0.2/T8 $\wedge$ 0.2/T8} = {0.2/T8}   {111*, 211*, 612*} = 0.2
{111*, 312*, 612*} = {0.4/T8 $\wedge$ 0.2/T8 $\wedge$ 0.2/T8} = {0.2/T8}   {111*, 211*, 612*} = 0.2

**Table- 14**

$$N_3^3$$

| 3-itemset | Min_Support |
|---|---|
| {111*, 211*, 312*} | 0.56 |
| {111*, 211*, 612*} | 0.2 |
| {111*, 211*, 612*} | 0.2 |

All 3-items considered for further process because their *support* >= $\beta_3^3$.

**4-Itemset**
{111*, 211*, 312*, 612*} = {0.4/T8 $\wedge$ 0.2/T8 $\wedge$ 0.2/T8} = {0.2/T8}
**Support of 4- Itemsets**                    {111*, 211*, 312*, 612*} = 0.2

**Table- 15**

$$N_4^3$$

| 4-itemset | Min_ Support |
|---|---|
| {111*, 211*, 312*, 612*} | 0.2 |

All 4-items considered for further process because their *support* >= $\beta_4^3$.

**Level k = 4**
Min_support ( $\beta_q^4$ ) = 0.16

**1-Itemset**                                   **Support of 1-Items**
{1111} = {0.2/T9, 0.33/T12}         {1111} = 0.53
{1112} = {0.4/T1, 0.16/T2, 0.16/T3}  {1112} = 0.72
{1113} = {0.33/T5, 0.4/T8}           {1113} = 0.73
{2112} = {0.4/T1, 0.16/T2, 0.2/T8}   {2112} = 0.76
{3122} = {0.2/T1, 0.33/T2, 0.2/T6, 0.2/T8}   {3122} = 0.93
{6123} = {0.2/T8, 0.4/T9, 0.6/T10}   {6123} = 1.2

**Table- 16**

$$N_1^4$$

| 1-itemset | Min_ Support |
|---|---|
| {1111} | 0.53 |
| {1112} | 0.72 |
| {1113} | 0.73 |
| {2112} | 0.76 |
| {3122} | 0.93 |
| {6123} | 1.2 |

All 1-items considered for further process because their *support* >= $\beta_1^4$.

**2-Itemset**                                   **Support of 2-Itemsets**
{1111, 1112} = { }                   {1111, 1112} = { }
{1111, 1113} = { }                   {1111, 1113} = { }
{1111, 2112} = { }                   {1111, 2112} = { }
{1111, 3122} = { }                   {1111, 3122} = { }
{1111, 6123} = {0.4/T9 $\wedge$ 0.2/T9} = 0.2   {1111, 6123} = 0.2
{1112, 1113} = { }                   {1112, 1113} = { }
{1112, 2112} = {0.4/T1 $\wedge$ 0.4/T1, 0.16/T2 $\wedge$ 0.16/T2} = {0.4/T1, 0.16/T2}   {1112, 2112} = 0.56
{1112, 3122} = {0.4/T1 $\wedge$ 0.2/T1, 0.16/T2 $\wedge$ 0.33/T2} = {0.2/T1, 0.16/T2}   {1112, 3122} = 0.36
{1112, 6123} = { }                   {1112, 6123} = { }
{1113, 2112} = {0.4/T8 $\wedge$ 0.2/T8} = {0.2/T8}   {1113, 2112} = 0.2
{1113, 3122} = {0.4/T8 $\wedge$ 0.2/T8} = {0.2/T8}   {1113, 3122} = 0.2
{1113, 6123} = {0.4/T8 $\wedge$ 0.2/T8} = {0.2/T8}   {1113, 6123} = 0.2
{2112, 3122} = {0.4/T1 $\wedge$ 0.2/T1, 0.16/T2 $\wedge$ 0.33/T2, 0.2/T8 $\wedge$ 0.2/T8} = {0.2/T1, 0.16/T2, 0.2/T8}   {2112, 3122} = 0.56
{2112, 6123} = {0.2/T8 $\wedge$ 0.2/T8} = {0.2/T8}   {2112, 6123} = 0.2
{3122, 6123} = {0.2/T8 $\wedge$ 0.2/T8} = {0.2/T8}   {3122, 6123} = 0.2

**Table- 17**

$$N_2^4$$

| 2-itemset | Min_ Support |
|---|---|
| {1111, 6123} | 0.2 |
| {1112, 2112} | 0.56 |
| {1112, 3122} | 0.36 |
| {1113, 2112} | 0.2 |
| {1113, 3122} | 0.2 |
| {1113, 6123} | 0.2 |
| {2112, 3122} | 0.56 |
| {2112, 6123} | 0.2 |
| {3122, 6123} | 0.2 |

The {1111, 1112}, {1111, 1113}, {1111, 2112}, {1111, 3122}, {1112, 1113}, {1112, 6123} **cannot be considered for further process because their** *support* $<= \beta_2^4$.

**3-Itemset**     **Support of 3- Itemsets**
{1112, 2112, 3122} = {0.4/T1 ∧ 0.4/T1    {1112, 2112, 3122} = 0.36
∧ 0.2/T1, 0.16/T2 ∧ 0.16/T2 ∧ 0.33/T2}
= {0.2/T1, 0.16/T2}
{1113, 2112, 3122} = {0.4/T8 ∧ 0.2/T8 ∧    {1113, 2112, 3122} = 0.2
0.2/T8} = {0.2/T8}
{1113, 2112, 6123} = {0.4/T8 ∧ 0.2/T8 ∧    {1113, 2112, 6123} = 0.2
0.2/T8} = {0.2/T8}
{1113, 3122, 6123} = {0.4/T8 ∧ 0.2/T8 ∧    {1113, 3122, 6123} = 0.2
0.2/T8} = {0.2/T8}
{2112, 3122, 6123} = {0.2/T8 ∧ 0.2/T8 ∧    {2112, 3122, 6123} = 0.2
0.2/T8} = {0.2/T8}

**Table- 18**

$N_3^4$

| 3-itemset | Min_ Support |
|---|---|
| {1112, 2112, 3122} | 0.36 |
| {1113, 2112, 3122} | 0.2 |
| {1113, 2112, 6123} | 0.2 |
| {1113, 3122, 6123} | 0.2 |
| {2112, 3122, 6123} | 0.2 |

All 3-items considered for further process because their *support* = $\beta_3^4$.

**4-Itemset**
{1113, 2112, 3122, 6123} = {0.4/T8 ∧ 0.2/T8 ∧ 0.2/T8, ∧ 0.2/T8}
**Support of 4- Itemsets**   {1113, 2112, 3122, 6123} = 0.2

**Table- 19**

$N_4^4$

| 4-itemset | Min_ Support |
|---|---|
| {1113, 2112, 3122, 6123} | 0.2 |

Step-8
Support of each q-itemset is calculated as given in the following results:
Step= 9
From the results as performed by Step-7 and 8, the sets of frequent 1-itemsets, 2-itemsets, 3-itemsets and 4-itemset at level 1are given in Table 4, 5 ,6, 7, 8 ,9 ,10 and11,12, respectively.

Step-10
This step is just for increment the value of k in which if $q > \chi$, then the process is going to Step-12.
Step-11
This step is looking for possible/candidate *p*-itemsets fromn*Nq-1* at level k as there is no more candidate itemsets (as p=4) then go to Step-12. Otherwise, the process is going to Step-5.
Step-12
The step is to calculate every confidence of each possible association rules as follows:
At level k =1

$$Conf(1*** \Rightarrow 2***) = \frac{Support(1***,2***)}{Support(1***)} = \frac{0.92}{2.89} = 0.31$$

.
.
.

Similarly the confidence calculated for itemset

$$Conf(1***\wedge 2*** \Rightarrow 3***) = \frac{Support(1***,2***,3***)}{Support(1***,2***)} = \frac{0.72}{0.92}$$
$$= 0.78$$

$$Conf(1***\wedge 2***\wedge 3*** \Rightarrow 6***) = \frac{Support(1***,2***,3***,6***)}{Support(1***,2***,3***)}$$
$$= \frac{0.36}{0.72} = 0.5$$

Similarly the Confidence calculated for levels k =2,3&4 are following:

$$Conf(11**\wedge 21** \Rightarrow 31**) = \frac{Support(11**,21**,31**)}{Support(11**,21**)}$$
$$= \frac{0.72}{0.92} = 0.78$$

$$Conf(11**\wedge 21**\wedge 31** \Rightarrow 61**) = \frac{Support(11**,21**,31**,61**)}{Support(11**,21**,31**)}$$
$$= \frac{0.36}{0.72} = 0.5$$

$$Conf(111*\wedge 211* \Rightarrow 312*) = \frac{Support(111*,211*,312*)}{Support(111*,211*)}$$
$$= \frac{0.56}{0.76} = 0.73$$

$$Conf(111*\wedge 211*\wedge 312* \Rightarrow 612*) = \frac{Support(111*,211*,312*,612*)}{Support(111*,211*,312*)}$$
$$= \frac{0.2}{0.56} = 0.35$$

$$Conf(1112\wedge 2112 \Rightarrow 3122) = \frac{Support(1112,2112,3122)}{Support(1112,2112)}$$
$$= \frac{0.36}{0.56} = 0.64$$

**Conclusion:**
This paper introduced a novel algorithm for generating fuzzy multilevel association rule. The algorithm is based on the concept that the larger number of items purchased in a transaction means the lower level of association among the items in the transaction. The algorithm works well with problems involving vagueness in data relationships, which are represented by fuzzy set concepts. The proposed fuzzy mining algo-rithm can thus generate large itemsets level by level and then derive fuzzy association rules from transaction dataset. Finally, an illustrated example is given to clearly demonstrate and understand steps of the algorithm.